\documentclass[prb,amsmath,amssymb,twocolumn]{revtex4}
\usepackage{t1enc}              
\usepackage[latin2]{inputenc}   

\usepackage{graphicx}
\usepackage{dcolumn}
\usepackage{bm}

\begin{document}


\title{Electron states in the quantum wire with periodic serial structure}

\author{Jarosław Kłos}
\email{klos@amu.edu.pl}

\affiliation{Surface~Physics~Division,~Faculty~of~Physics,~Adam
Mickiewicz~University,\\ ul.Umultowska 85, 61-614 Pozna\'{n},
Poland}

\date{\today}

\begin{abstract}
A model quantum wire embedded in a matrix permeable to electron
waves is investigated in terms of electronic states. The wire is
assumed to have a 1D crystal structure. Through electron waves
propagating in its surroundings, lateral modes are coupled with
Bloch waves propagating along the wire axis, which results in
modes splitting into multiplets. The results presented in this
study have been obtained by direct solution of the Schrödinger
equation in the effective mass approximation.
\end{abstract}


\maketitle

\section{Introduction}

Having become the subject of intensive research \cite{crook,
gilbert, minchul, belluci}, electron transport in quantum wires
includes an interesting problem of transport in wires showing
structural periodicity, which can be due to material composition
or geometry variations \cite{dobrzyn}. A periodic quantum wire can
be regarded as a quasi one-dimensional crystal; transmission
through it depends on the crystal band structure as well as on
open lateral modes.

A more general treatment of the problem to be dealt with here can
be found in a paper by Kohn \cite{kohn} and in later studies
\cite{freeman, soven, jepsen}, in which electronic states in
crystalline thin films (composed of just a few atomic layers) are
investigated by means of the Green's function formalism. In the
1990s, in connection with the possibility of fabricating complex
semiconductor heterostructures with molecular beam epitaxy (MBE)
technology, papers on electronic properties of quantum wires began
to appear. A relatively low percentage of these studies deals with
periodic quantum wires, though \cite{persson,dress,piraux}. In
particular, the problem of electronic state existence in a
periodic quantum wire embedded in a permeable medium has not yet
been studied thoroughly enough.

The model of quantum wire considered in this paper shows
structural periodicity due to variations in material composition,
and reflected by periodic variations of the effective mass, as
well as of the effective potential felt by electrons moving along
the wire axis. The potential confining the electron motion to the
wire axis is assumed to be of finite value; consequently, the
electron waves do not form nodes on the wire borders and thus can
penetrate into the matrix in which the wire is embedded.

The following is implied by the periodicity of the wire structure
and the finite potential value in the matrix:
\begin{itemize}
  \item   The electron motion in space cannot be regarded as a
  superposition of two independent motions, one along the wire axis,
  the other perpendicular to it. This is why modes localized in the
   wire form multiplets. Each multiplet is determined by the number
   of nodes of the envelope function, $\Psi(y)$, in the wire, and each mode
   in a multiplet corresponds to a different Fourier component of
    Bloch wave $\Psi(x)$.
  \item    In spite of the finite value of the potential limiting the
  electron motion perpendicular to the wire axis, the number of multiplets
  is infinite. This is due to the reduction of the wave vector (referring
  to the electron motion along the wire axis) to the first Brillouin zone,
  which means that even for high energy values corresponding bound states
  exist in the quantum wire.
\end{itemize}

The discussed model is based on the effective mass approximation,
which, though not applicable when the electron energy is far from
the conduction band bottom level, through simplifications made,
allows a clear description of conditions of mode existence in a
periodic quantum wire.

\section{Model}

\begin{figure}
\includegraphics[width=7.5cm]{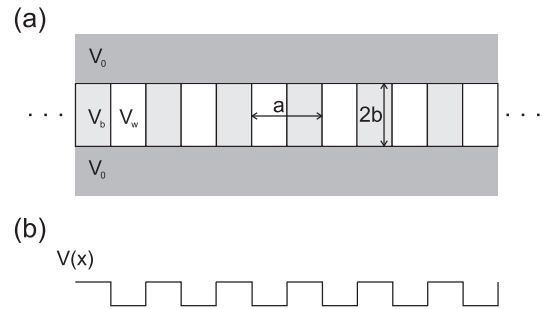}
\caption{(a) A schematic representation of a periodic quantum wire
composed of alternating segments of $Al_{x}GaAs_{1-x}$ (potential
barriers) and $GaAs$ (potential wells). The wire is embedded in an
$AlAs$ matrix. (b) The conduction band bottom level along the wire
axis.}\label{fig:f1}
\end{figure}

The system to be considered is schematically depicted in
Fig.~\ref{fig:f1}. A periodic quantum wire can be regarded as a
superlattice with two surfaces close to each other and
perpendicular to its layers. The wire is composed of alternating
segments of $GaAs$ and $Al_{x}GaAs_{1-x}$, and embedded in an
$AlAs$ matrix.

The electron motion with energy values close to the conduction
band bottom is described by the effective mass equation:
\begin{equation}
      \left[\frac{\hbar^{2}}{2} \nabla \frac{1}{m^{*}({\bm r})}\nabla+E-V({\bm r})\right]\Psi({\bm r})=0.   \label{eq:r1}
\end{equation}

Since the system is homogeneous along the $z$-axis, the electron
motion described in (\ref{eq:r1}) can be regarded as a
superposition of two motions, one taking place in the $x,y$-plane,
the other being a free motion along the $z$-axis. If the latter
motion is neglected ($k_{z}=0$), the system can be regarded as
two-dimensional.

A further factorization of $\Psi(x,y)$ is possible within the
quantum wire: $\Psi_{I}(x,y)=\Psi_{I}(x)\Psi_{I}(x)$, with the
following differential equations to be satisfied by components
$\Psi_{I}(x)$ and $\Psi_{I}(y)$:
\begin{eqnarray}
\left[\frac{\hbar^{2}}{2m_{0}} \frac{\partial}{\partial x}
\frac{1}{M(x)} \frac{\partial}{\partial x}
+E-V(x)-\frac{1}{M(x)}\lambda \right]\Psi_{I}(x)=0, \label{eq:r2}
\\
\left[\frac{\hbar^{2}}{2m_{0}} \frac{\partial ^2}{\partial y^{2}}+
\lambda \right]\Psi_{I}(y)=0, \label{eq:r3}
\end{eqnarray}
where $M(x)=m^{*}(x)/m_{0}$, and $\lambda$ is the separation
constant. The solution, $\Psi_{I}(x,y)$, has the following form:
\begin{equation}
\Psi_{I}(x,y)=u(x)e^{i x q}e^{i\kappa y},\label{eq:r4}
\end{equation}
where $u(x)$ denotes the periodic factor of the Bloch function,
and $\kappa=\sqrt{\frac{2m_{0}}{\hbar}\lambda}$.

Throughout the system, a solution of (\ref{eq:r1}) must fulfill
the conditions of continuity at the border between the quantum
wire $'I'$ and the matrix $'II'$:
\begin{eqnarray}
\left.\frac{\partial\Psi_{I}(x,y)} {\partial x}\right|_{y=\pm b
}&=&\left.\frac{\partial\Psi_{II}(x,y)}{\partial x}\right|_{y=\pm
b },\label{eq:r5}\\ \frac{1}{M(x)}
\left.\frac{\partial\Psi_{I}(x,y)} {\partial y}\right|_{y=\pm b
}&=&\frac{1}{M}\left.\frac{\partial\Psi_{II}(x,y)}{\partial
y}\right|_{y=\pm b },\label{eq:r6}
\end{eqnarray}
where $M(x)$ and $M$ denote the relative effective mass in the
wire and beyond it, respectively.

For condition (\ref{eq:r5}) to be fulfilled by the wave function
beyond the wire, $\Psi_{II}(x,y)$, for any $x$ value, the function
must have the following form:
\begin{equation}
\Psi_{II}(x,y)=\sum_{n=-\infty}^{\infty}c_{n}e^{i(q+Q_{n})x}f_{n}(y),
\label{eq:r7}
\end{equation}
$c_{n}$ denoting Fourier coefficients of the periodic factor of
the Bloch function, $\Psi_{I}(x)$.
\begin{equation}
u(x)=\sum_{n=-\infty}^{\infty}c_{n}e^{iQ_{n}x}. \label{eq:r8}
\end{equation}
The fulfillment of (\ref{eq:r5}) for any $x$ implies the following
boundary conditions for $f_{n}(y)$:
\begin{eqnarray}
f^{+}_{n}(y)\left.\right|_{y\rightarrow
\infty}=f^{-}_{n}(y)\left.\right|_{y\rightarrow \infty}=0.
\label{eq:r9}
\end{eqnarray}
The boundary condition:
\begin{eqnarray}
f^{+}_{n}(b)=f^{-}_{n}(-b)=1\label{eq:r10}
\end{eqnarray}
means a requirement of mode localization at the wire axis.
Superscripts $'+'$ and $'-'$ in (\ref{eq:r9}) and (\ref{eq:r9})
refer to the solutions in two half planes, defined by $y>b$ and
$y<-b$, respectively.

By including (\ref{eq:r7}) into (\ref{eq:r1}), we get:
\begin{eqnarray}
\sum_{n=-\infty}^{\infty}c_{n}e^{i(q+Q_{n})x}& \times & \nonumber
 \\
 \left[\frac{\partial^{2}}{\partial y^{2}
}-(q+Q_{n})^{2}+\frac{2 M}{\hbar^{2}}(E-V_{0})\right]f_{n}(y)&
\equiv & 0, \label{eq:r11}
\end{eqnarray}
where $V_{0}$ is the conduction band bottom level in the matrix.
This leads to the following system of differential equations for
$f_{n}(y)$ with boundary conditions (\ref{eq:r9}) and
(\ref{eq:r10}):
\begin{equation}
\left[\frac{\partial^{2}}{\partial y^{2} }-(q+Q_{n})^{2}+\frac{2
M}{\hbar^{2}}(E-V_{0})\right]f_{n}(y)=0.\label{eq:r12}
\end{equation}
The solution of (\ref{eq:r12}) has the form:
\begin{equation}
f^{\pm}_{n}(z)=e^{\mu_{n}(z\mp b)},\label{eq:r13}
\end{equation}
where
\begin{equation}
\mu_{n}=\sqrt{(q+Q_{n})^{2}+\frac{2M}{\hbar^{2}}(V_{0}-E)}.\label{eq:r14}
\end{equation}

A general solution of (\ref{eq:r3}) is a linear combination of
functions: $C_{1}e^{-\kappa y}+C_{2}e^{\kappa y}$. The symmetry of
the system allows independent matching of modes that are symmetric
or antisymmetric with respect to the wire axis. Thus, the solution
of (\ref{eq:r6}) can be found by independent mode matching at one
surface only ($y=b$):
\begin{eqnarray}
\frac{1}{M}\kappa \tan(\kappa
b)=\frac{1}{M(x)}\frac{\sum_{n}\mu_{n}c_{n}e^{(q+Q_{n})x}}{\sum_{n}c_{n}e^{(p+Q_{n})x}},\label{eq:r15}\\
\frac{1}{M}\kappa \cot(\kappa
b)=-\frac{1}{M(x)}\frac{\sum_{n}\mu_{n}c_{n}e^{(q+Q_{n})x}}{\sum_{n}c_{n}e^{(p+Q_{n})x}}.\label{eq:r16}
\end{eqnarray}
Equations (\ref{eq:r15}) and (\ref{eq:r16}) refer to symmetric and
antisymmetric modes, respectively.

In the vicinity of the conduction band bottom the effective mass
is a linear function of the band bottom level, $V(x)$:
\begin{equation}
M(x)=A+B V(x).\label{eq:r17}
\end{equation}
Being a periodic function ($V(x+na)=V(x)$), potential $V(x)$ can
be Fourier-expanded:
\begin{equation}
V(x)=\sum_{n=-\infty}^{\infty}v_{n}e^{i Q_{n} x}.\label{eq:r18}
\end{equation}
With only the most significant component of the sum in
(\ref{eq:r18}) taken into account, $M(x)$ is approximately
represented by its mean value:
\begin{equation}
M(x)\approx\overline{M(x)}=A+Bv_{0}.\label{eq:r19}
\end{equation}
By including this approximation into (\ref{eq:r15}) and
(\ref{eq:r16}), we get the following form of the two equations:
\begin{eqnarray}
\sum_{n=-\infty}^{\infty}c_{n}e^{i(q+Q_{n})x}\left(M\mu_{n}-\overline{M(x)}\kappa\tan(\kappa
b )\right)=0,\label{eq:r20}\\
\sum_{n=-\infty}^{\infty}c_{n}e^{i(q+Q_{n})x}\left(M\mu_{n}+\overline{M(x)}\kappa\cot(\kappa
b )\right)=0.\label{eq:r21}
\end{eqnarray}
to be satisfied for any $x$ (i.e. throughout the length of the
border between the wire and the matrix). This condition is
fulfilled when:
\begin{eqnarray}
M\mu_{n}-\overline{M(x)}\kappa\tan(\kappa b )=0,\label{eq:r22}\\
M\mu_{n}+\overline{M(x)}\kappa\cot(\kappa b )=0.\label{eq:r23}
\end{eqnarray}

The values of $\kappa$ corresponding to known energy values $E$
can be found from (\ref{eq:r22}) and (\ref{eq:r23}). In the limit
case, when the matrix is impermeable to electron waves
($V_{0}\longrightarrow\infty$), $\kappa$ takes values
$\kappa_{l}=l\pi/(2b)$; odd $l$ values correspond to modes that
are symmetric with respect to the wire axis, and even $l$ values
to antisymmetric ones. Then, wave propagation along the wire axis
takes place in  effective potential:
\begin{equation}
U(x)=V(x)+\frac{2m^{*}(x)}{\hbar}l\pi/(2b).\label{eq:r24}
\end{equation}
When the potential in the matrix is finite, each $n$ corresponds
to a mode multiplet. Each mode in a multiplet corresponds to a
different Fourier component of electron wave $\Psi(x)$, and
involves the fulfillment of condition:
\begin{equation}
(q+Q_{n})^{2}+\frac{2M}{\hbar^{2}}(V_{0}-E)>0.\label{eq:r25}
\end{equation}
Note that modes corresponding to the lowest Fourier components
($n=0,\pm1,\pm2,\ldots$) will be successively eliminated from
multiplets with increasing energy. As a consequence of
(\ref{eq:r14}), (\ref{eq:r22}) and (\ref{eq:r23}), the limit value
of $\kappa$ for $|n|\rightarrow\infty$ is $l\pi/(2b)$ in each
multiplet.

\section{Results}

\begin{figure}
\includegraphics[width=7.5cm]{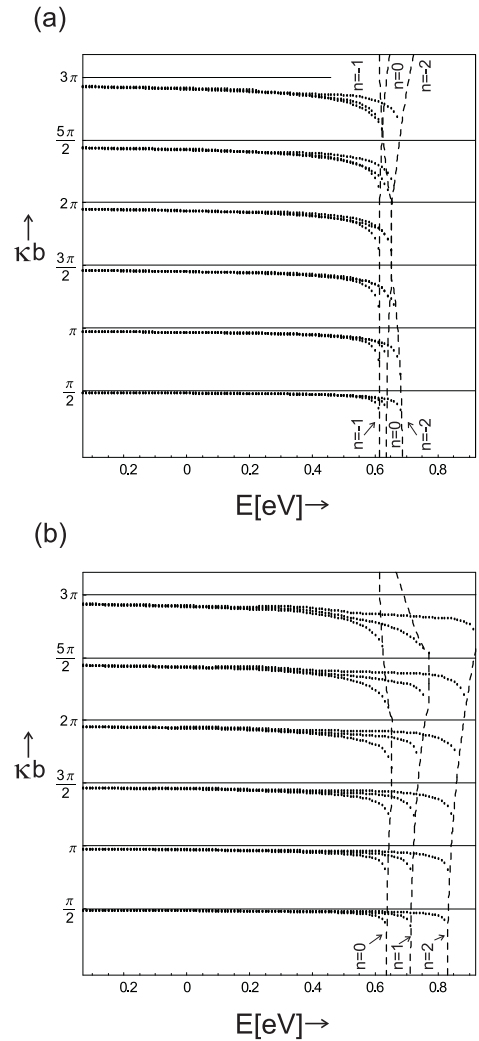}
\caption{Parameter $\kappa_{l,n}$ plotted versus energy for modes
(a) $n=0,-1,-2$ and (b) $n=0,1,2$. Dots represent calculation
points; horizontal lines indicate values approached by
$\kappa_{l,n}$ when $n\rightarrow\pm\infty$ for the first six
multiplets, $l=1,2,\ldots,5$; dashed lines indicate cut-off energy
values. }\label{fig:f2}
\end{figure}

The calculations performed refer to a quantum wire composed of
alternating $GaAs$ and $Al_{0.35}GaAs_{0.65}$ segments,
representing potential wells and barriers, respectively. The wire
is embedded in a uniform homogeneous $AlAs$ matrix. The energy
reference level ($E=0$) is assumed to coincide with the conduction
band bottom in an $Al_{0.35}GaAs_{0.65}$ barrier. The wells and
barriers are assumed to be of equal width,
$40\mathop{A}\limits^{o}$. The width of the quantum wire is
assumed to be constant throughout its length, and to equal
$200\mathop{A}\limits^{o}$.

The values of $\kappa_{l,n}$ corresponding to localized modes with
different energy values have been found from (\ref{eq:r22}) and
(\ref{eq:r23}). Subscript $l$ refers to the number of $\Psi(y)$
nodes in the quantum wire ($l=1,2,3,\ldots$ corresponds to
$0,1,2,\ldots$ nodes, respectively). Subscript $n$ refers to the
$n$-th Fourier component of Bloch function $\Psi(x)$.

\begin{figure}
\includegraphics[width=7.5cm]{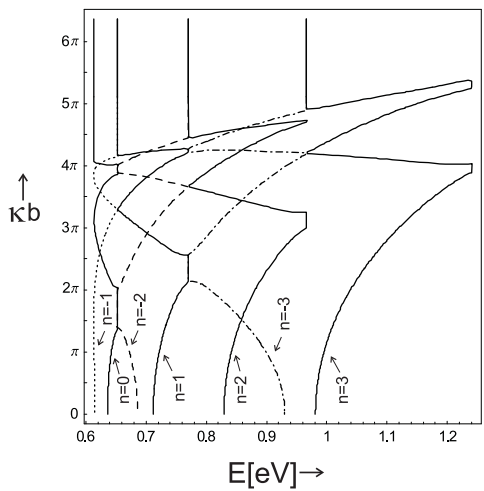}
\caption{Parameter $\kappa$ plotted versus cut-off energy for
different modes. Each curve corresponds to a different mode
number, $n$. }\label{fig:f3}
\end{figure}

Fig.~\ref{fig:f2} shows $\kappa_{l,m}$ plotted against electron
energy. For clarity reasons, $\kappa_{l,m}$ values are expressed
in units of $1/b$, $b$ denoting the half-width of the wire. Values
$\kappa b = \pi/2, 3/2 \pi, 5/2\pi,\ldots$ ($\kappa b =\pi, 2\pi,
3\pi,\ldots$) correspond to symmetric (antisymmetric) closed-end
modes (with nodes at the wire borders), and to
$n\rightarrow\pm\infty$. The calculations have been performed for
the first five Fourier components $n=0, \pm1, \pm2$. The
$\kappa_{l,n}$ values corresponding to modes in the same multiplet
(i.e. having the same number of nodes, or $l$) begin to differ
significantly at higher electron energy values. The splitting is
also found to grow in extent with increasing $l$. The dashed lines
in the plot represent the 'cut-off energy' values, at which, in
accordance with (\ref{eq:r25}), modes corresponding to successive
Fourier components ($n=0, \pm1, \pm2, \pm3, \ldots$) are
eliminated as the electron energy increases. In Fig.~\ref{fig:f3},
$\kappa$ is plotted versus cut-off energy for different modes
(different $n$ values). For high $\kappa$ values $q(\kappa,E)=0$,
and according to (\ref{eq:r25}), the cut-off energy for modes only
differing in sign of $n$ is identical. The variations of cut-off
energy for lower $\kappa$ values are a consequence of the
variability of $q(\kappa,E)$.

\begin{figure}
\includegraphics[width=7.5cm]{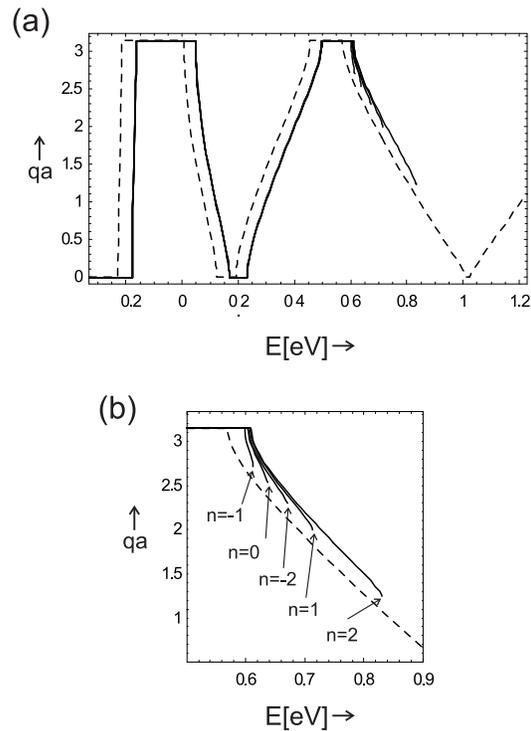}
\caption{Dispersion curves for the second (l=2) multiplet of
modes. Plotted for reference, the respective dispersion curve for
an infinite superlattice ($b\rightarrow\infty$) is represented by
the dashed line. (b) The area of elimination of modes $n=0,\pm
1,\pm 2$ shown in close-up. }\label{fig:f4}
\end{figure}

According to (\ref{eq:r2}) and (\ref{eq:r4}), $\kappa_{l,n}$
defines the effective potential for an electron moving along the
wire in mode $(l,m)$:
\begin{equation}
U(x)=V(x)+\frac{\hbar^2}{2m^{*}(x)}\kappa_{l,n} b.\label{eq:r26}
\end{equation}
At low electron energy values $\kappa_{l,m}$ remains constant.
Therefore, the dispersion relation, $q(E,\kappa_{l,n})$ (for the
motion along the wire), is shifted along the energy axis, the
extent of the shift depending on $\kappa$. Only at higher energy
values, in the range in which mode elimination occurs, the shape
of the dispersion curve is found to be significantly changed.
Fig.~\ref{fig:f4} shows the dispersion relation,
$q(E,\kappa_{2,n})$, for the second multiplet. The solid lines
represent dispersion curves for the lowest modes ($n=0, \pm1,
\pm2, \pm3$) in the multiplet. The dashed lines represent the
dispersion curve for an infinite semiconductor superlattice
($b\rightarrow\infty$, $\kappa\rightarrow0$).

\section{Conclusions}

Localized electronic modes in a periodic quantum wire
($\Psi(x,y)\rightarrow 0$ for $y\rightarrow\pm\infty$) are defined
by two quantum numbers: (1) $l$, referring to the number of nodes
of the envelope function in the direction perpendicular to the
wire axis; and (2) $n$, denoting the number of the corresponding
Fourier component of the wave propagating along the wire.

The dispersion curve, $q(E,\kappa_{l,m})$, corresponding to each
mode multiplet (fixed $l$) is shifted along the energy axis
towards higher energy values with respect to the dispersion curve
corresponding to an infinite superlattice ($b\rightarrow\infty$).
The relative shift of the $q(E,\kappa_{l,n})$ curve corresponding
to individual modes (differing in $n$) in a multiplet remains
minor as long as the electron energy is low (i.e. the left-hand
side of (\ref{eq:r25}) is much greater than zero); not until the
energy gets close to the cut-off value does the mode splitting
become noticeable.




\end{document}